# Restriction-induced time-dependent transcytolemmal water exchange: Revisiting the Kärger exchange model

Diwei Shi[1], Fan Liu[2], Sisi Li[2], Li Chen[1], Xiaoyu Jiang[3,4], John C. Gore[3,4,5,6], Quanshui Zheng[1], Hua Guo[2], and Junzhong Xu[3,4,5,6*]

[1] Center for Nano and Micro Mechanics, Department of Engineering Mechanics, Tsinghua University, Beijing, China

[2] Center for Biomedical Imaging Research, Department of Biomedical Engineering, School of Medicine, Tsinghua University, Beijing, China

[3] Institute of Imaging Science, Vanderbilt University Medical Center, Nashville, Tennessee.

[4] Department of Radiology and Radiological Sciences, Vanderbilt University Medical Center, Nashville, Tennessee.

[5] Department of Biomedical Engineering, Vanderbilt University, Nashville, Tennessee.

[6] Department of Physics and Astronomy, Vanderbilt University, Nashville, Tennessee.

\* **Corresponding author**: Address: Institute of Imaging Science, Department of Radiology and Radiological Sciences, Vanderbilt University Medical Center, 1161 21st Avenue South, AA 1105 MCN, Nashville, TN 37232-2310, United States. Fax: +1 615 322 0734.

E-mail address: junzhong.xu@vanderbilt.edu (Junzhong Xu)

# Abstract

The Kärger model and its derivatives have been widely used to incorporate transcytolemmal water exchange rate, an essential characteristic of living cells, into analyses of diffusion MRI (dMRI) signals from tissues. The Kärger model consists of two homogeneous exchanging components coupled by an exchange rate constant and assumes measurements are made with sufficiently long diffusion time and slow water exchange. Despite successful applications, it remains unclear whether these assumptions are generally valid for practical dMRI sequences and biological tissues. In particular, barrier-induced restrictions to diffusion produce inhomogeneous magnetization distributions in relatively large-sized compartments such as cancer cells, violating the above assumptions. The effects of this inhomogeneity are usually overlooked. We performed computer simulations to quantify how restriction effects, which in images produce edge enhancements at compartment boundaries, influence different variants of the Kärger-model. The results show that the edge enhancement effect will produce larger, time-dependent estimates of exchange rates in e.g., tumors with relatively large cell sizes (>10 μm), resulting in overestimations of water exchange as previously reported. Moreover, stronger diffusion gradients, longer diffusion gradient durations, and larger cell sizes, all cause more pronounced edge enhancement effects. This helps us to better understand the feasibility of the Kärger model in estimating water exchange in different tissue types and provides useful guidance on signal acquisition methods that may mitigate the edge enhancement effect. This work also indicates the need to correct the overestimated transcytolemmal water exchange rates obtained assuming the Kärger-model.

# Introduction

Biological tissues contain water in a range of different microenvironments, such as cells, axons, neurons, myelin sheaths, blood vessels, and various tissue interstitial spaces. Water molecules cross cell membranes, either directly through the lipid bilayer or facilitated by water channel proteins known as aquaporins (AQP) (1,2). The overall exchange rate determines the cell membrane permeability to water (3,4), which is an important characteristic of living cells (5) and has been a parameter of interest in cancer (6-8) and brain (9-11) studies. Variations in transcytolemmal water exchange may be associated with cell volume regulation (12), tumor aggressiveness (13,14), treatment response (15-17), brain edema (11,18), and various neurological disorders, such as Parkinson's disease (19). In addition, cell membrane ion pump activity, a measure of mitochondrial metabolism, has been shown to also affect water exchange, which suggests that the transcytolemmal water exchange rate may be a sensitive indicator of cellular energy turnover (17,20,21). Therefore, it would be valuable to develop noninvasive imaging techniques to quantify transcytolemmal water exchange in vivo. Such techniques could not only provide useful biomarkers for monitoring alterations in cellular properties, especially membrane permeability but also have the potential to improve the specificity of related pathological diagnoses. Previous imaging studies have been implemented based on compartmental relaxation rates modified by adding paramagnetic contrast agents (22,23). However, such methods suffer from limits on the contrast agent distribution achievable in living tissues (24) in addition to possible side effects of the agent. Therefore, it is of interest to develop a contrast agent-free imaging method.

Diffusion MRI (dMRI) provides a non-invasive, contrast agent-free approach to probe in vivo biological tissues. Some dMRI-based methods have been developed to characterize specific microstructures on the cellular and subcellular levels (~10 μm) based on the time dependence of restricted water diffusion (25,26). In general, these methods assume that dMRI signals arise from different compartments (muti-compartment) to construct biophysical models (27-31), but methods such as VERDICT (27) and IMPLUSED (28,29) usually ignore water exchange between compartments. This simplification results not only in the loss of important information about transcytolemmal water exchange but also leads to significant biases in the estimation of other microstructural parameters, especially an underestimation of intracellular volume fraction (32). Therefore, it is critical to incorporate water exchange into dMRI-based quantitative imaging to provide richer and more accurate information on microstructural features in clinical applications.

To tackle the above problem, Kärger proposed a theoretical framework to describe the evolution of magnetizations from two compartments in which water diffuses with Gaussian behaviors and undergoes inter-compartmental exchange, dubbed the Kärger model (33,34). Price et al. extended this model by introducing the effect of restricted diffusion in spherical pores (i.e., a geometric model of cells) (35), motivating the study of transcytolemmal water exchange based on the Kärger model. Several Kärger-model-based methods have been developed to quantify water exchange in dMRI. Specifically, the constant diffusion gradient (CG) method was proposed to measure the transcytolemmal water exchange rate constant $k_{in}$ by keeping the diffusion gradient amplitude $g$ and duration $\delta$ constant while changing the diffusion gradient separation $\Delta$ (36,37). The FEXI (filtered-exchange imaging) method employs double diffusion encoding imaging sequences (DDE) to measure apparent water exchange rates (AXR) (4,38,39). The NEXI (Neurite Exchange Imaging) method was built on an anisotropic Kärger model for a coherent fiber tract, using dMRI signals acquired at multiple diffusion weightings (i.e., $b$ values) and diffusion times (40). It can derive microstructural parameters of nerve fibers including the water exchange time $t_{ex}$. In a concurrent study, a similar model is termed as the "Standard Model" with exchange between neurites and extracellular water (SMEX)(41). Other models such as MADI (24) and CEXI (42) have also been developed to tackle exchange. In addition, Jiang et al have combined the Kärger model and the IMPLUSED method to incorporate water exchange into dMRI-based cell size imaging (43). Cell size and transcytolemmal water exchange rate can be estimated simultaneously by using the Kärger model to calculate the signals acquired at a long diffusion time ($t_{diff} > 30$ ms and The above Kärger-model-based methods have been successfully implemented in preclinical and clinical research, such as in vivo rat cortex (40), breast cancer (43,44), and intracranial brain tumors (9).

However, the Kärger model is based on three assumptions (45):
1) The diffusion time $t_{diff}$ is sufficiently long, i.e., $t_{diff} \gg l^2/D$, where $l$ and $D$ are the compartmental size and diffusivity, respectively. (long diffusion time)
2) The transcytolemmal water exchange rate constant $k_{in}$ is relatively slow, which means the exchange time $\tau_{in}$ $(= 1/k_{in})$ is also long, i.e., $\tau_{in} \gg l^2/D$. (slow water exchange)

3) The gradient strength $q = \gamma\delta g$ is relatively small, i.e., $q \ll 1/l_c$, where $l_c$ is the coarse-grained length associated with the size, shape, and arrangement of the compartments. (small diffusion-weighted)

These conditions limit the feasibility of the Kärger model in some cases with large compartment sizes, short diffusion times, fast water exchange, or high b-values, especially relevant in tumor tissues with large cell diameters (46) and high membrane permeability (47). When the above conditions are not met, particularly condition (1), the magnetization distribution is likely inhomogeneous inside one compartment, such as in large cells. This is similar to the restriction-induced inhomogeneous magnetization distribution as reported previously (48-52), i.e., "edge enhancement", which is also known as "localization regime" effect at high gradients (53-55): because the diffusion of molecules near an boundary (whether impermeable or semi-permeable) is more restricted, their local transverse magnetization attenuates more slowly compared to the bulk magnetization. Such a phenomenon will be more pronounced with shorter diffusion times, such as OGSE acquisitions with moderately high gradient frequencies (56), and then it becomes inappropriate to oversimplify a biological system into two uniform signal components associated with intra- and extracellular compartments. Previous studies typically focused on the role of the "edge enhancement" or "localization regime" effect on the asymptotic behavior of dMRI signals(55), while its impact on the measurement of water exchange using the Kärger based models remain unclear. Recently, some works reported overestimation of water exchange rate constants and cell membrane permeability in large cancer cells using numerical simulations and in vivo experiments (32,57), which may be caused by the cell membrane restriction on molecular diffusion. Therefore, this motivated us to investigate how such an edge enhancement effect influences the estimation of transcytolemmal water exchange rate in the Kärger model-based methods.

In this work, computer simulations were performed to quantitatively investigate the impact of the magnetization inhomogeneity on the estimated $k_{in}$ based on the Kärger model. In addition, we also systematically explore how dMRI pulse sequences (e.g., different b values and diffusion times) and microstructural features (e.g., cell size and membrane permeability) affect the measurement of $k_{in}$ in current commonly-used water exchange imaging methods (CG, FEXI, NEXI/SMEX) and a recently proposed microstructural model (JOINT) (43). The results shed light on the influences of sequence and microstructural parameters on the estimation of transcytolemmal water exchange, and the necessity to correct the Kärger-model-derived exchange rate constants in cancer studies.

## Materials and Methods

### Theory

Considering water exchange occurs between two compartments, the steady state concentrations (= number of molecules per volume) of compartmental water molecules can be expressed as:

$$\begin{cases} \dfrac{dC_{in}}{dt} = -k_{in}C_{in} + k_{ex}C_{ex} \\ \dfrac{dC_{ex}}{dt} = -k_{ex}C_{ex} + k_{in}C_{in} \end{cases}, \quad (1)$$

where $C_{in}$ and $C_{ex}$ are the concentrations of water molecules in the intra- and extracellular compartments, respectively. $k_{in}$ is the water exchange rate constant of water molecules from intra- to extracellular space, and $k_{ex}$ is from extra- to intracellular space. At equilibrium, $C_{in}$ and $C_{ex}$ are constants and $v_{in}k_{in} = v_{ex}k_{ex}$, where $v_{in}$ and $v_{ex}$ are intra- and extracellular water fractions, respectively, and $v_{in} + v_{ex} = 1$. Note that Eq. (1) describes the biophysical process of transcytolemmal water exchange without any involvement of magnetic resonance.

Based on Eq. (1), the Kärger model was developed to describe both diffusion and magnetization exchange between intra- and extracellular compartments in dMRI experiments, namely,

$$\begin{cases} \dfrac{dM_{in}}{dt} = -q^2 D_{in}M_{in} - k_{in}^m M_{in} + k_{ex}^m M_{ex} \\ \dfrac{dM_{ex}}{dt} = -q^2 D_{ex}M_{ex} - k_{ex}^m M_{ex} + k_{in}^m M_{in} \end{cases}, \quad (2)$$

where $M_{in}$ and $M_{ex}$ are total magnetizations in the intra- and extracellular compartments, respectively, $D_{in}$ and $D_{ex}$ are the compartmental diffusivities, and $k_{in}^m$ and $k_{ex}^m$ are intra-to-extra and extra-to-intra exchange rate constants of magnetizations. The traditional Kärger model assumes $k_{in}^m$ and $k_{ex}^m$ (i) are time-independent (i.e., constants) and (ii) equal to the exchange rate constants $k_{in}$ and $k_{ex}$ of water molecules, respectively. This assumption has been widely used in dMRI-based quantitative imaging methods incorporating water exchange (4,36-40,43).

However, Eqs. (1) and (2) describe the transcytolemmal exchange of water molecules and magnetizations, respectively, so their derived metrics are not necessarily the same, i.e., $k_{in}^m$ and $k_{ex}^m$ may not equal to $k_{in}$ and $k_{ex}$, respectively. This is particularly true when the magnetization distribution is inhomogeneous due to restricted diffusion near compartment boundaries. Typically, the magnetizations close to the boundaries (e.g. cell membranes) are higher than those far away from the boundaries before well mixing, especially for the intracellular compartment with restricted diffusion. On the other hand, most water molecules involved in transcytolemmal water exchange also come from the neighborhood of the cell membranes. Therefore, the net outflow ratio (net outflow amount/ total amount) of the magnetizations should be usually higher than that of the molecules, i.e., $k_{in}^m$ and $k_{ex}^m$ usually exhibit larger values than $k_{in}$ and $k_{ex}$, respectively, due to the restriction-induced edge enhancement effect. In addition, the magnetization distribution within the compartment is time-varying during the application of a gradient waveform, and then $k_{in}^m$ and $k_{ex}^m$ should be defined as time-dependent. A more generalized Kärger model is therefore given as:

$$\begin{cases} \dfrac{dM_{in}}{dt} = -q^2 D_{in} M_{in} - k_{in}^m(t) M_{in} + k_{ex}^m(t) M_{ex} \\ \dfrac{dM_{ex}}{dt} = -q^2 D_{ex} M_{ex} - k_{ex}^m(t) M_{ex} + k_{in}^m(t) M_{in} \end{cases}. \quad (3)$$

To characterize the time-dependent deviation of $k_{in}^m(t)$ or $k_{ex}^m(t)$ from the constant $k_{in}$ or $k_{ex}$, we introduce two dimensionless parameters:

$$\begin{cases} \alpha_{in}(t) = \dfrac{k_{in}^m(t)}{k_{in}} \\ \alpha_{ex}(t) = \dfrac{k_{ex}^m(t)}{k_{ex}} \end{cases}. \quad (4)$$

To summarize the deviation during the whole gradient waveform, we define two effective parameters $\overline{\alpha_{in}}$ and $\overline{\alpha_{ex}}$ by integrating over time, which are the time-averaged ratios of $k_{in}^m$ to $k_{in}$, and $k_{ex}^m$ to $k_{ex}$:

$$\begin{cases} \overline{\alpha_{in}} = \dfrac{1}{TE} \displaystyle\int_0^{TE} \alpha_{in}(t) dt \\ \overline{\alpha_{ex}} = \dfrac{1}{TE} \displaystyle\int_0^{TE} \alpha_{ex}(t) dt \end{cases}. \quad (5)$$

With the parameters defined in Eqs. (4) and (5), we are able to quantify the influence of restriction-induced edge enhancement effects on the estimation of transcytolemmal water exchange with various diffusion MRI sequences and microstructural properties.

**Simulation**

A finite difference method (58) was used to simulate the time evolution of the magnetization distribution in a one-dimensional, two-compartmental, exchanging model system. This tissue model is periodic along the x-axis, i.e., each compartment is exchanging with other compartments at both ends, as shown in Figure 1. Note that the actual extracellular compartments are usually narrower than the intracellular ones, we set up two different compartments with unequal lengths, and the wide compartment represents the intracellular space while the narrow one represents the extracellular space. Therefore, in what follows, the subscripts "in" and "ex" of the parameters are replaced by "w" (wide) and "n" (narrow), respectively.

The microstructural features were set up as follows: the wide and narrow compartmental sizes, $d_w$ and $d_n$, were 15 and 5μm, resulting in an "intracellular" (i.e. wide compartment) volume fraction of 75%; both $D_w$ and $D_n$, were 1.56 μm²/ms (29); The membrane permeability $P_m$, set as 0.02μm/ms, was used to control the water exchange rate in the simulations, instead of $k_w$ (water exchange rate constant of molecules from wide to narrow compartment). A larger $P_m$ indicates more frequent water exchange. Specifically, the relationship between $P_m$ and $k_w$ is given by (57):

$$\frac{1}{P_m} = \frac{6}{k_w \cdot d_w} - \frac{d_w}{10 D_w}. \tag{6}$$

For the calculation of time-dependent exchange rate constants of magnetizations, as shown in Figure 1, we count the net outflow magnetizations from the wide and narrow compartments after each time step, respectively. Then $k_w^m(t)$ and $k_n^m(t)$ can be obtained by dividing the net outflow magnetizations by the product of the time step $\Delta t$ with the total amount within the compartments, and they were calculated at each time point during the simulations. Note that $k_w$ and $k_n$ were constants, which correspond to the exchange of matter (water molecules), we only calculated their values after the first-time step. Then, the time-dependent dimensionless parameters $\alpha_w(t)$, $\alpha_n(t)$, and their time-averaged values ($\overline{\alpha_w}$ and $\overline{\alpha_n}$) can be obtained by Eq. (4) and (5).

**Studies**

Two studies were performed.
1) We quantified the influence of restriction-induced edge enhancement effects on different dMRI sequences, including PGSE, double PGSE, and oscillating gradient spin-echo (OGSE). Specifically, the time evolution of magnetization distribution across the modeled tissue was shown to demonstrate the edge enhancement effect. Moreover, $\alpha_w(t)$ and $\alpha_n(t)$ were shown to demonstrate the time dependence of the edge enhancement and its influences in the intra-and extracellular spaces. The corresponding acquisition parameters are shown in Table 1.

2) We investigated the impact of the edge enhancement on the estimation of transcytolemmal water exchange in several Kärger-model-based methods that measure water exchange, including CG, NEXI/SMEX, FEXI, and JOINT methods. Specifically, $\overline{\alpha_w}$ and $\overline{\alpha_n}$ were calculated to quantify the influences of the edge enhancement effect with different diffusion pulse sequence parameters (gradient separation $\Delta$, duration $\delta$, and amplitude g) and microstructural features (wide compartment size $d_w$, membrane permeability $P_m$, compartmental diffusivities $D_w$ and $D_n$). The respective variables and their corresponding ranges are shown in Table 2.

Table 1 The acquisition parameters of different dMRI sequences in study (1)

| Sequences | $\delta$ (ms) | $\Delta$ (ms) | TE (ms) | $t_m$(ms) | g (mT/m) |
|---|---|---|---|---|---|
| PGSE | 4 | 40 | 55 | / | 250 |
| Double PGSE | 6 | 30 | 55 | 100 | 180 |
| OGSE | 40 | 50 | 100 | / | 75 |

*For the double PGSE sequence, all parameters are the same for the two independent PGSE sequences (same below). *$t_m$ is the mixing time in the double PGSE sequence.

Table 2 (a) the acquisition parameters of different Kärger-model-based methods and their corresponding ranges in study (2).

| Method | Sequences | $\delta$ (ms) | $\Delta$ (ms) | $t_m$(ms) | g (mT/m) | b (ms/μm²) | N | $t_{diff}$(ms) |
|---|---|---|---|---|---|---|---|---|
| CG | PGSE | 10~30 | 80~2000 | / | 10~90 | / | / | / |
| NEXI /SMEX | PGSE | 2~7 | 10~60 | / | 10~300 | 0~8 | / | / |
| FEXI | Double PGSE | 6 | 15~40 | 10~3000 | / | 0~5 | / | / |
| JOINT | PGSE | 9 | 43,53,73 | / | / | 0~3 | / | 20~100 |
| OGSE | 40 | 50 | / | / | 0~3 | 1,2,3 | 2~18 |

Table 2 (b) The microstructural features and their corresponding ranges in study#2

| Microstructural feature | $d_w$(μm) | $d_n$(μm) | $D_w$(μm²/ms) | $D_n$(μm²/ms) | $P_m$(μm/ms) |
|---|---|---|---|---|---|
| Range | 5~20 | 5 | 1~3 | 1~3 | 0.001~0.1 |

*We investigated the impact of acquisition and microstructural parameters on $\overline{\alpha_w}$ and $\overline{\alpha_n}$ by the control variables approach, the detailed settings of variables are shown in the captions of Figure 3~6. *The above acquisition parameters or ranges are obtained from previously published clinical and experimental work. Note that in the NEXI/SMEX study, we performed the acquisition protocol of NEXI due to its lower gradient strength requirements, which is easier to implement on dMRI equipment.

In addition, we introduced a series of characteristic lengths and used the ratios of the compartment sizes to them as the x-coordinates, as shown in the **Results** section, to demonstrate more intuitive results and conclusions:

1) Effective diffusion length: $l_{diff} = \sqrt{2 \cdot D \cdot t_{diff}}$
2) Gradient characteristic length: $l_g = \left(\frac{D}{\gamma g}\right)^{1/3}$
3) B-value characteristic length: $l_b = \sqrt{TE/b}$
4) Permeability characteristic length: $l_{Pm} = D/P_m$
5) Mixed characteristic length: $l_{tm} = \sqrt{2 \cdot D \cdot t_m}$

where $t_{diff}$ is the effective diffusion time, which equals to $(\Delta - \delta/3)$ for PGSE and $\delta/4N$ for OGSE; $D$ usually equals to $D_w$, except for the case of a variation in $D_n$, where it equals to $D_n$; TE is the overall duration of the sequence.

## Results

Figure 2 visualizes the amplitude of the magnetization distribution in both intra- and extracellular spaces with the PGSE (a), double PGSE (b), and OGSE (c) sequences. The magnetization is particularly inhomogeneous in the larger intracellular space, while close to homogeneous in the narrow extracellular space. This is because the compartmental barriers restrict the surrounding water diffusion, resulting in less diffusion-induced local magnetization amplitude attenuations. This effect occurs close to barriers and gradually propagates away from barriers via diffusion, i.e., intra-compartmental "mixing". In the narrow extracellular space, most water molecules are influenced by barriers so the magnetization is close to homogeneous. Note that this restriction-induced inhomogeneous distribution of magnetizations, i.e., "edge enhancement" or "localization regime", was reported previously(48,49,51,55).

Figure 3 shows the time-dependent ratios $\alpha_w(t)$ and $\alpha_n(t)$ for the PGSE (a), double PGSE (b), and OGSE (c) sequences. The value of $\alpha_w(t)$ is typically larger than 1 while $\alpha_n(t)$ is always close to 1. Such an overestimation of the water exchange rate is consistent with the results in Figure 2. At any moment, transcytolemmal water exchange mostly involves molecules close to barriers. These "edge" molecules carry larger local magnetizations than the molecules far away from barriers. Then the percentage of exchanged magnetizations should be higher than that of exchanged water molecules, which results in a higher exchange rate constant of magnetizations than that of molecules, i.e., $k_w^m(t) > k_w$, as shown in Figure 3. Moreover, such overestimation of the exchange rate gradually diminishes after the diffusion gradient is off, presumably due to intra-compartmental water mixing that reduces the inhomogeneous distribution of magnetization. In contrast, the magnetization distribution within the narrow compartment is almost uniform for all three sequences because most molecules encounter barriers similarly in this compartment.

Figure 4 lists the variation of $\overline{\alpha_w}$ with individual pulse sequence (Δ and g) or microstructural parameters ($d_w$, $D_w$, $D_n$, and $P_m$) in the CG method. Most $\overline{\alpha_w}$ values are larger than 1, and the variations are obvious within the set ranges of the Δ, g, $d_w$, and $D_w$. These four factors play dominant roles for $\overline{\alpha_w} > 1$, (i.e., $k_w^m > k_w$ in the time-averaged sense during the signal acquisition). In contrast, the impact of narrow compartment diffusivity $D_n$ and membrane permeability $P_m$ is more limited. Especially for $D_n$, $\overline{\alpha_w}$ is almost invariant as it increases. In summary, the phenomenon of $k_w^m > k_w$ caused by the edge enhancement effect is usually non-negligible when we use the CG method to measure water exchange among tissues with large microstructural sizes ($d_w > 12$μm, note that $d_w = 12$μm corresponds to the gray dashed line in Figure 4 (c)), such as typical cancer cells. Furthermore, the results in Figure 4 also demonstrate that decreasing the gradient amplitude $g$ (larger $l_g$) and lengthening Δ (larger $l_{diff}$) can effectively reduce the impact of the edge enhancement effect, making the actual exchange rate constant of magnetizations closer to that of molecules in the CG method. The corresponding results for $\overline{\alpha_w}$ (same below) are shown in Supplement Martials, where the $\overline{\alpha_w}$ values are all near 1.

Figure 5 lists the variation of $\overline{\alpha_w}$ with each variable (*b*-value, $g$, $d_{iw}$, $D_w$, $D_n$, and $P_m$) in the NEXI/SMEX method. Similar to the results for CG, most $\overline{\alpha_w}$ values exceed 1, and three factors, *b*-value, $g$, and $d_w$ play dominant roles for $\overline{\alpha_w} > 1$. The impact of $D_w$ is relatively weak, and the other two factors, $D_n$ and $P_m$, have almost no impact. When the NEXI/SMEX method is implemented to measure water exchange between intra- and extra-axonal compartments (axon diameter < 6µm), the restriction-induced edge enhancement effect can be ignored due to small microstructural sizes. However, if this method is used in tissues with larger-sized cells (> 10µm, $d_w = $ 10µm corresponds to the colorful vertical dashed lines in Figure 5 (c)), such as typical cancer cells (59), or other larger compartments, such as soma (cell diameter ~11µm in the ex-vivo mouse gray matter(30) and ~18µm in the in-vivo rat cortex and hippocampus(40)), it is necessary to correct the overestimation of $k_{in}$ (i.e., $k_w$ in this work) caused by the edge enhancement effect. In addition, decreasing the overall diffusion weighting $b = \gamma^2 \delta^2 g^2 (\Delta - \delta/3)$ (larger $l_b$) and lengthening $\Delta$ (larger $l_{diff}$) with a fixed *b*-value can reduce the deviation of $k_w^m$ from $k_w$, i.e., $\overline{\alpha_w} \to 1$ in the NEXI/SMEX method.

Figure 6 lists the variation of $\overline{\alpha_w}$ with different parameters (*b*-value, $t_m$, $d_w$, $D_w$, $D_n$, and $P_m$) in the FEXI method. Similar to the results shown in Figures 4 and 5, most $\overline{\alpha_w} > 1$, and the *b*-value and $d_w$ play dominant roles. The impact of the mixing time $t_m$ and wide compartment diffusivity $D_w$ is relatively weak but noticeable. The impact of $D_n$ and $P_m$ is still limited and may be ignored. Similarly, the barrier-induced edge enhancement effect cannot be ignored in the implementation of the FEXI method to measure water exchange within tissues with larger microstructural sizes (> 10µm, $d_w = $ 10µm corresponds to the colorful vertical dashed lines in Figure 6 (c)), such as tumors. Furthermore, decreasing *b*-values (larger $l_b$), lengthening $t_m$ (larger $l_{tm}$) and $\Delta$ (larger $l_{diff}$) with a fixed *b* can derive $k_w^m$ more consistent with $k_w$ in the FEXI method.

Figure 7 lists the variation of $\overline{\alpha_w}$ with individual parameters (*b*-value, $t_{diff}$, $d_w$, $D_w$, $D_n$, and $P_m$) in the JOINT method. The number of periods $N = 1, 2, 3$ in the trapezoidal cosine-modulated OGSE, representing $t_{diff}$= 10, 5, and 3.33 ms for $\delta = $ 40ms, and $t_{diff} = 40, 50, 70$ ms in the PGSE. Similar to the results of other sets of simulation experiments (Figure 4-6), most $\overline{\alpha_w}$ values are larger than 1. The effective diffusion time $t_{diff}$, *b*-value, and wide compartment size $d_w$ play dominant roles.

The impact of $D_w$ is relatively weak for the PGSE sequences and limited for OGSE. The other two factors, $D_n$ and $P_m$, still have almost no impact. For the implementation of JOINT to measure transcytolemmal water exchange in tumors (cell size $> 10\mu m$, $d_w = 10\mu m$ corresponds to the colorful vertical dashed lines in Figure 7 (c)), the edge enhancement-induced $k_w^m > k_w$ is non-negligible and needs to be corrected to improve the overestimation of the transcytolemmal water exchange rate. In addition, decreasing *b*-values (larger $l_b$) and lengthening $t_{diff}$ (larger $l_{diff}$, using a longer Δ for PGSE and a longer δ for OGSE) can reduce the impact of the edge enhancement effect and make $k_w^m \to k_w$. However, lengthening $t_{diff}$, especially for the OGSE sequences, will reduce the sensitivity of time-dependence-based dMRI methods to microstructural features. Simultaneously, using lower *b*-values will reduce the accuracy and robustness of the model fitting. Therefore, there is a trade-off between the accurate estimations of microstructural sizes (e.g., cancer cell size) and water exchange rates when designing the acquisition protocol for JOINT.

## Discussion

In recent years, several dMRI methods have been developed to non-invasively measure transcytolemmal water exchange in vivo, based on the classic Kärger-model. However, three main assumptions, i.e., a sufficient long diffusion time $t_{diff}$ ($\gg l^2/D$), a slow water exchange rate constant $k_{in}$ ($\ll D/l^2$), and a small gradient strength $q$ ($\ll 1/l_c$), may not be valid in biological tissues with relatively large microstructural sizes and high membrane permeability, such as tumors. Under this circumstance, the membrane-induced restricted diffusion inside cells leads to an inhomogeneous distribution of magnetizations, with higher magnetizations close to the boundaries, i.e., cell membranes. Currently, this edge enhancement effect is overlooked, although it may cause an overestimation of the transcytolemmal water exchange rate constant in practical applications. In this work, we performed a series of computer simulations to quantitatively and systematically investigate the deviation of the actual exchange rate constant of magnetizations from that of water molecules, which is caused by restriction-induced magnetization inhomogeneity. With an increasing interest in mapping transcytolemmal water exchange in cancer studies, this work sheds light on extracting more accurate biological information through Kärger-model-based methods.

As shown in Figures 4~7, in the implementation of the four water exchange imaging methods included in this work, the phenomenon of $k_w^m > k_w$ caused by the edge enhancement becomes non-negligible for large compartmental sizes (typically for $d_w > 10$ μm). For the central nervous system imaging in the brain, this effect seems not an issue due to small axon or cell sizes (< 6μm) (60,61). However, for tumor microstructural imaging, although it is usually not a concern in the extracellular irregular and narrow interstitial space, the restriction-induced edge enhancement effect needs to be considered in the intracellular compartment due to large cancer cell sizes (10~30 μm) (46,62). In summary, it should be kept in mind that the Kärger-model-derived "$k_{in}$" is the actual exchange rate constant of magnetizations $k_{in}^m$, which is usually larger than $k_{in}$, the true exchange rate constant of water molecules, in any tissues with large cell sizes (> 10 μm) when incorporating the Kärger-model.

In addition to the microstructural features, the parameters of dMRI sequence also show an obvious effect on $\overline{\alpha_w}$, the time-averaged ratio of the magnetization and molecule exchange rate constants. For the effective gradient strength $q = \gamma\delta g$ of the PGSE sequences, it is intuitive that $\overline{\alpha_w}$ becomes larger as $q$ increases, whether by increasing the diffusion gradient $g$ or lengthening the duration $\delta$. However, for the gradient separation $\Delta$, the dependency of $\overline{\alpha_w}$ on $\Delta$ seems non-monotonic when $q$ is fixed (note that not a fixed $b$-value). As shown in Figure 4 (a), $\overline{\alpha_w}$ decreases with a longer $\Delta$ (corresponds to a smaller $d_w/l_{diff}$) in the CG methods. In contrast, as shown in Figure 5 (c), $\overline{\alpha_w}$ increases with a longer $\Delta$ for $\Delta \leq 45$ ms in the NEXI/SMEX methods. Here, we further investigate this non-monotonic relationship and show the results in Figure 8, where the overall duration TE of the PGSE sequence is set to (a): TE = $\Delta + \delta + 10$ms and (b): TE = constant ($> \Delta + \delta$), respectively. Regardless of how to set TE, $\overline{\alpha_w}$ first becomes larger and then decreases with an increasing $\Delta$ under a fixed $q$ (the value of $\delta g$ is fixed as 1200 mT·ms/m for each curve), but the transition points, i.e., the $\Delta$ values corresponding to the maximum $\overline{\alpha_w}$, are usually different on the "$\overline{\alpha_w}$ v.s. $\Delta$" curves. Specifically, the maximum $\overline{\alpha_w}$ occurs at $\Delta \approx 20$ms (corresponds to the gray vertical dashed lines) for the situations of TE = $\Delta + \delta + 10$ms; Then for the other set of situations of TE =constant, the transition point becomes larger as TE increases within the range of 25~55ms (corresponds to the gray vertical area). Furthermore, the results in Figure 8 (a) show that lengthening $\delta$ (i.e. decreasing $g$) can reduce the impact of the edge enhancement effect and make $\overline{\alpha_w} \to 1$ when $q$ is fixed. Similarly, lengthening TE can also reduce the deviation of $k_w^m$ from $k_w$, as shown in Figure 8 (b). In summary, the results in Figures 4~8 provide some useful guidance on signal acquisition to mitigate the overestimation of the

transcytolemmal water exchange rate constant, that is:

1) try to use a longer $\Delta$ in the PGSE sequence to increase the *b*-value, instead of increasing the gradient strength $q = \gamma \delta g$ (i.e. larger $l_{diff}$ and smaller $d_w/l_{diff}$).
2) try to use a longer $\delta$ and decrease g under a fixed $q$ value (i.e. larger $l_g$ and smaller $d_w/l_g$).
3) try to lengthen the overall duration TE of the gradient sequence (i.e. larger $l_b$ and smaller $d_w/l_b$).

The restriction-induced edge enhancement effect results in inhomogeneous magnetization distribution in large compartments, which violates the assumption of two homogenous, exchanging components in the Kärger model. This is the essential reason responsible for the overestimation of the transcytolemmal water exchange rate using the Kärger model when the edge enhancement effect is pronounced. Intuitive thought is to split the compartment with inhomogeneous magnetization distribution into two or more so that the Kärger model includes three or more exchanging components. However, it is well-known challenging to fit a system with multiple exchanging components (63-65). It is expected challenging for the Kärger model to contain three or more exchanging components. It is practical to keep two exchanging components particularly for in vivo studies, although the edge enhancement effect may overestimate water exchange.

One major limitation of this work is that we have performed relevant simulations in 1D cases. This is because, under such circumstances, it is straightforward to visualize the restriction-induced edge enhancement effect and easier to understand the underlying biophysical mechanism. The main conclusion, however, can be extended to 3D qualitatively and explain the overestimations of the transcytolemmal water exchange rate constant observed in practical applications. The same approach was also used in previous studies(48,55,66). Future work includes the translation of 1D simulations to realistic histology-informed 3D models (24,58,67). Another limitation is that the analytical forms of $\overline{\alpha_w}$ or $k_w^m(t)$ cannot be derived here because the edge enhancement effect depends on different pulse sequence parameters and complex microstructural features. In future work, it is plausible to quantify the influence of the edge enhancement effect in specific Kärger-model-based methods and develop corrections to estimate the transcytolemmal water exchange rate.

This work included only three different diffusion gradient sequences (PGSE, double PGSE, and OGSE) and four dMRI-based water exchange imaging methods (CG,

NEXI/SMEX, FEXI, and JOINT), but the investigations on the restriction-induced edge enhancement effect and the corresponding conclusions can be extended to other sequences(68) and methods. Finally, we emphasize that all Kärger-model-based methods should incorporate the impact of inhomogeneous magnetization distributions and edge enhancements within biological tissues with larger compartmental sizes under diffusion gradients with finite durations.

## Conclusion

The influences of restriction-induced edge enhancement effect on the estimation of transcytolemmal water exchange using the Kärger model was comprehensively investigated using computer simulations in three different diffusion pulse sequences (PGSE, double PGSE, and OGSE) and four Kärger model-based methods that quantify water exchange (i.e. CG, NEXI/SMEX, FEXI, and JOINT). The results suggest that the previously overlooked edge enhancement effect leads to an overestimation of transcytolemmal water exchange rate constant in e.g., cancerous tissues with relatively large cell sizes (>10 μm), but should have negligible effects in e.g. the brain with relatively small cell and axon sizes (< 6 μm). Moreover, stronger diffusion gradients, longer diffusion gradient durations, and larger cell sizes all lead to more severe overestimations of water exchange. These results assist in better understanding how the Kärger model works differently in various tissues, but also point to the necessity to correct estimates of transcytolemmal water exchange obtained using Kärger-model-based methods, particularly in tumors.

## Acknowledgments

The authors thank Drs. Adam Anderson and Daniel F. Gochberg for stimulating discussion.## Reference

**Figures**

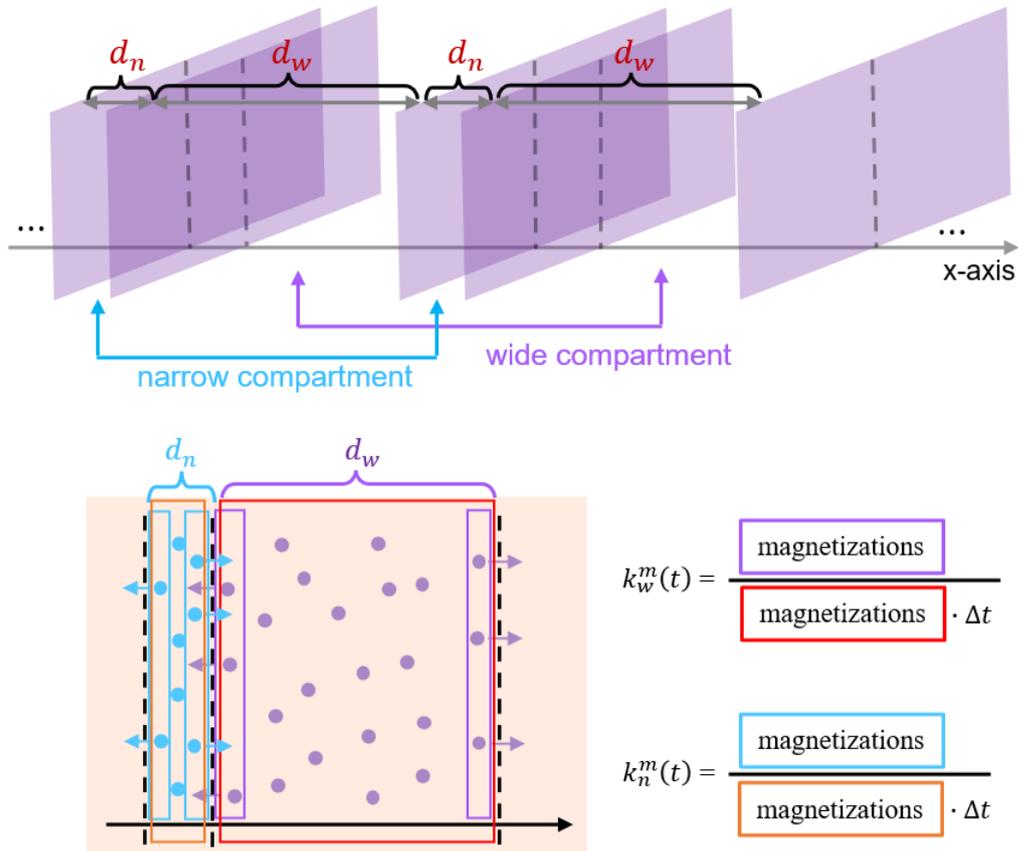

Figure 1. Schematic diagrams of the structure with parallel planes that was used in the finite difference simulation (top) and the calculation of time-dependent exchange rate constants of magnetizations (bottom). Interleaved wide and narrow compartments mimics intra and extracellular spaces, respectively. The diffusion gradient is perpendicular to the planes. The time-dependent exchange rate constants $k_w^m(t)$ and $k_n^m(t)$ are defined as the net outflow magnetizations divided by the compartmental total magnetization multiplied by the time step $\Delta t$.

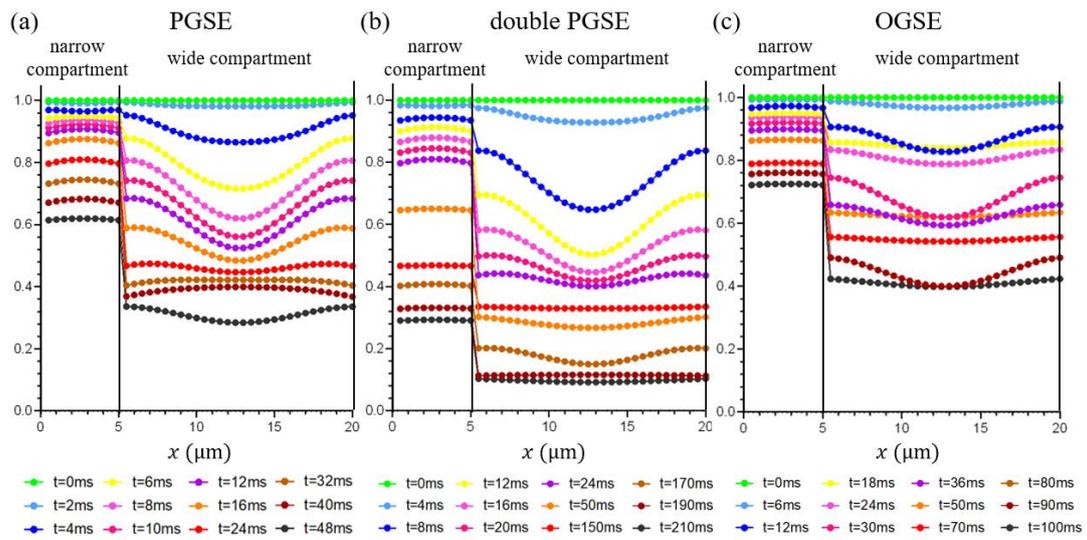

Figure 2. The amplitude of the magnetization distribution in both wide and narrow spaces with the PGSE (a), double PGSE (b), and OGSE (c) sequences. The different colors of the dots and lines represent different time points. In overall, the magnetizations in space are decaying with time. But for the wide compartment, the magnetization near the boundaries is usually higher than that in the compartment center, while the magnetization distribution within the narrow compartment is almost uniform.

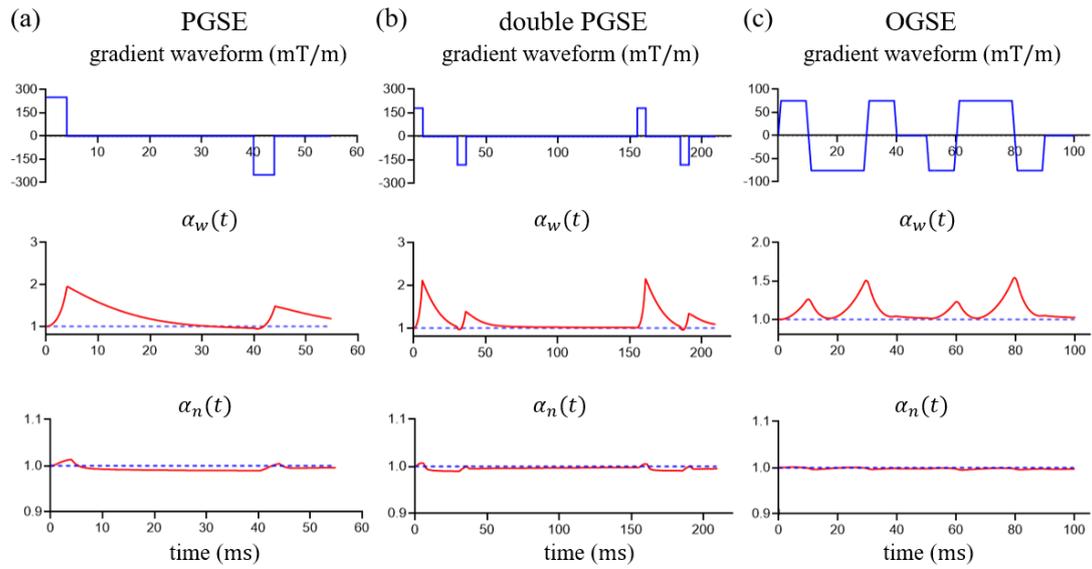

Figure 3. The time-dependent ratios $\alpha_w(t)$ and $\alpha_n(t)$ for the PGSE (a), double PGSE (b), and OGSE (c) sequences. The value of $\alpha_w(t)$ is usually larger than 1 while $\alpha_n(t)$ is always close to 1.

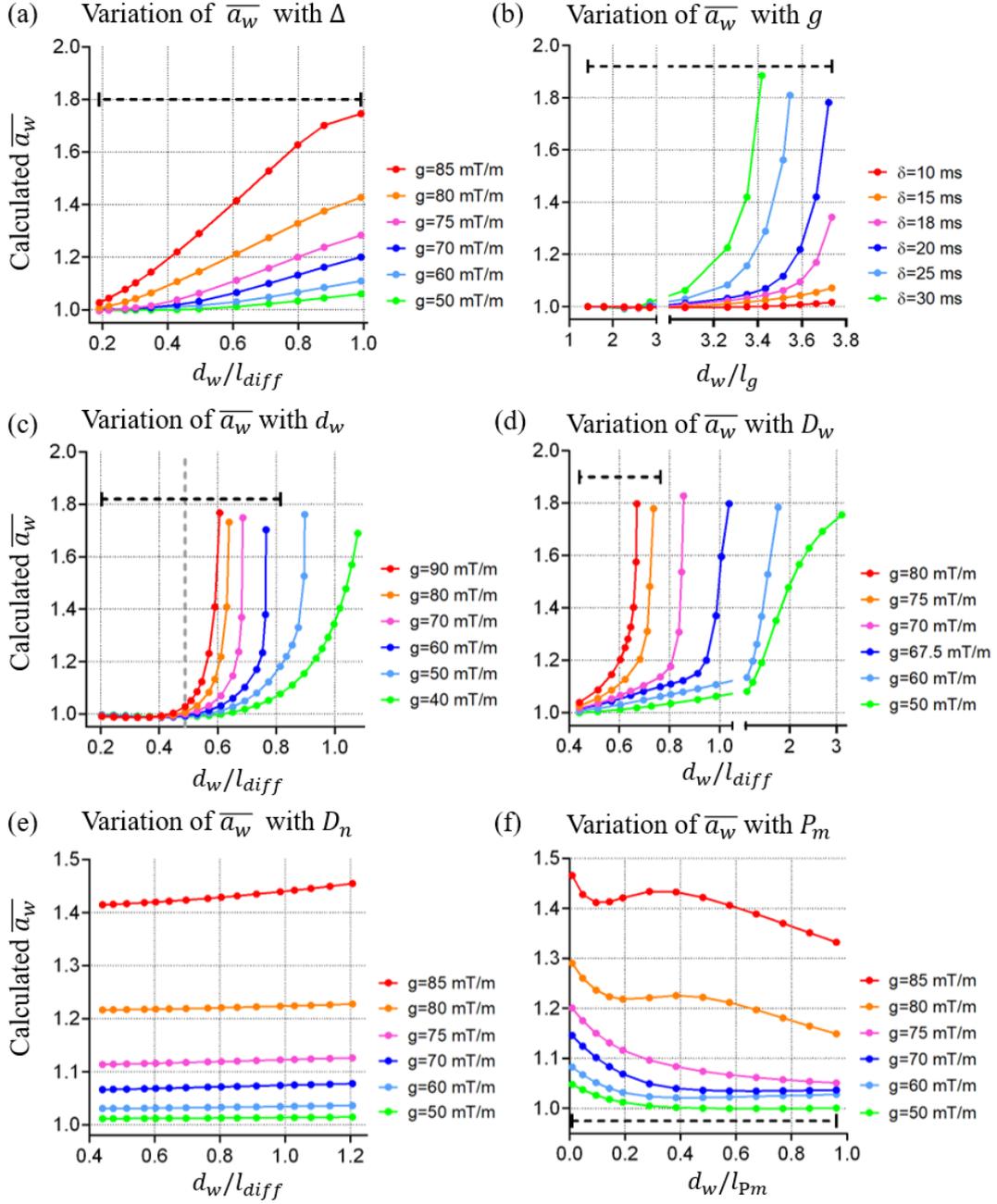

Figure 4. The calculated time-averaged ratio $\overline{a_w}$ as a function of the dimensionless ratio in the CG method. The basic settings in this group of simulations are: $\delta = 20$ ms, $\Delta = 200$ ms, $d_w = 15$ μm, $d_n = 5$ μm, $D_w = D_n = 1.56$ μm$^2$/ms, $P_m = 0.02$ μm/ms. (a). The variation of $\overline{a_w}$ with $\Delta$. The different colors of the dots and lines represent different $g$ settings, and the other variables are the same as the basic settings (same in the subfigures (c)-(f)). The black horizontal dashed line indicates that $\Delta$ varies from 1500~80ms (from left to right, same below). (b). The variation of $\overline{a_w}$ with $g$. Here the different colors of the dots and lines represent different $\delta$ settings, and the other variables are the same as the basic settings. The black horizontal dashed line

indicates that $g$ varies from 10~90mT/m. (c). The variation of $\overline{a_w}$ with $d_w$. The black horizontal dashed line indicates that $d_w$ varies from 5~20μm. (d). The variation of $\overline{a_w}$ with $D_w$. The black horizontal dashed line indicates that $D_w$ varies from 1~3μm²/ms. (e). The variation of $\overline{a_w}$ with $D_n$. (f). The variation of $\overline{a_w}$ with $P_m$. The black horizontal dashed line indicates that $P_m$ varies from 0.001~0.1μm/ms.

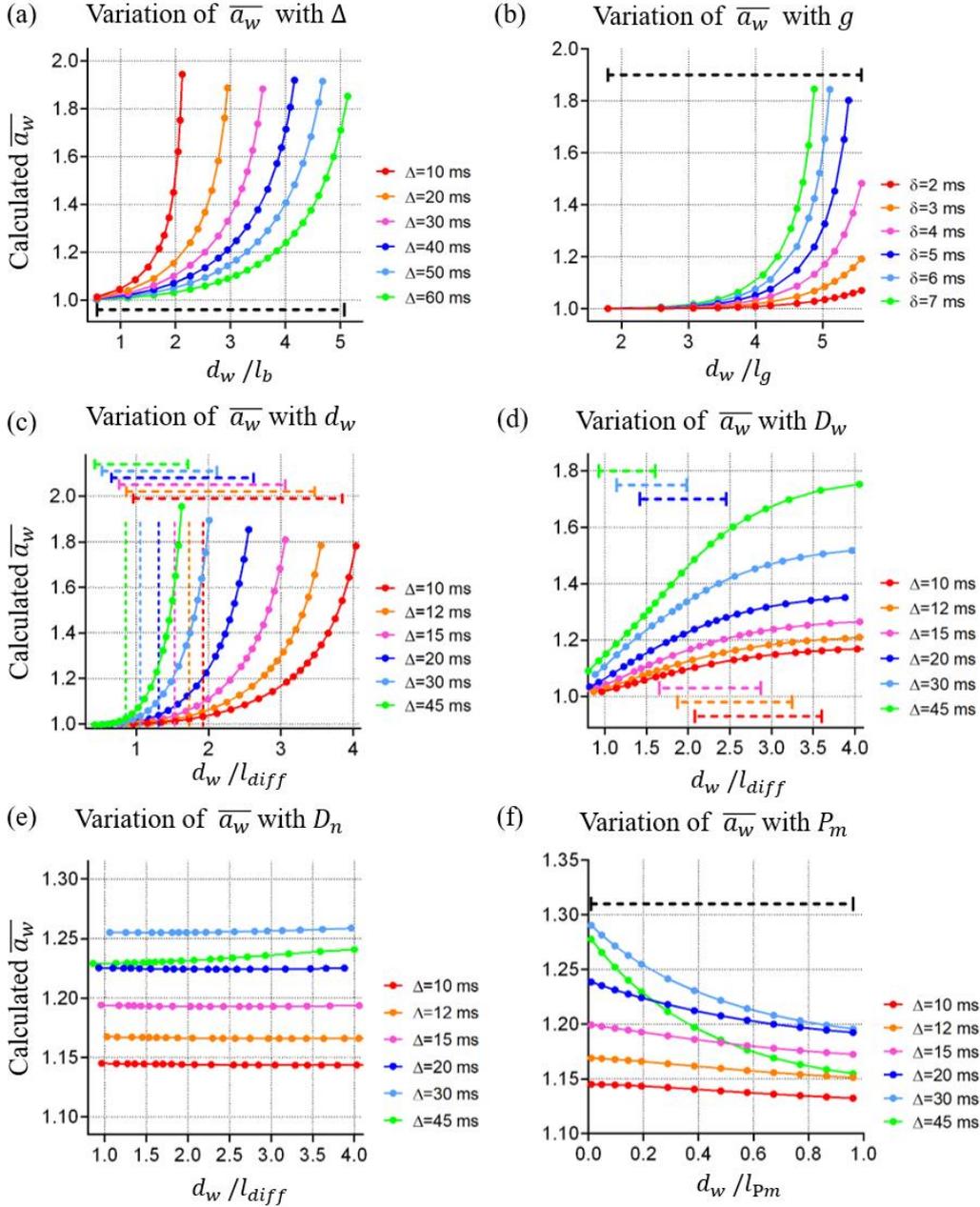

Figure 5. The calculated time-averaged ratio $\overline{a_w}$ as a function of the dimensionless ratio in the NEXI/SMEX method. The basic settings in this group of simulations are: $\delta = 4$ ms, $\Delta = 40$ ms, $g = 250$ mT/m, $d_w = 15$μm, $d_n = 5$μm, $D_w = D_n = 1.56$μm$^2$/ms, $P_m = 0.02$μm/ms. (a). The variation of $\overline{a_w}$ with the *b*-value (adjust $g$ by changing $g$). The different colors of the dots and lines represent different $\Delta$ settings, and the other variables are the same as the basic settings (same in the subfigures (c)-(f)). The black horizontal dashed line indicates that *b* varies from 0~8 ms/μm$^2$ (from left to right, same below). (b). The variation of $\overline{a_w}$ with $g$. Here the different colors of the dots and lines represent different $\delta$ settings, and the other variables are the same as the basic settings. The black horizontal dashed line indicates that $g$ varies from 10~300 mT/m. (c). The variation of $\overline{a_w}$ with $d_w$. The colorful

horizontal dashed line indicates that $d_w$ varies from 5~20μm. (d). The variation of $\overline{a_w}$ with $D_w$. The colorful horizontal dashed line indicates that $D_w$ varies from 1~3μm²/ms. (e). The variation of $\overline{a_w}$ with $D_n$. (f). The variation of $\overline{a_w}$ with $P_m$. The black horizontal dashed line indicates that $P_m$ varies from 0.001~0.1μm/ms.

# FEXI

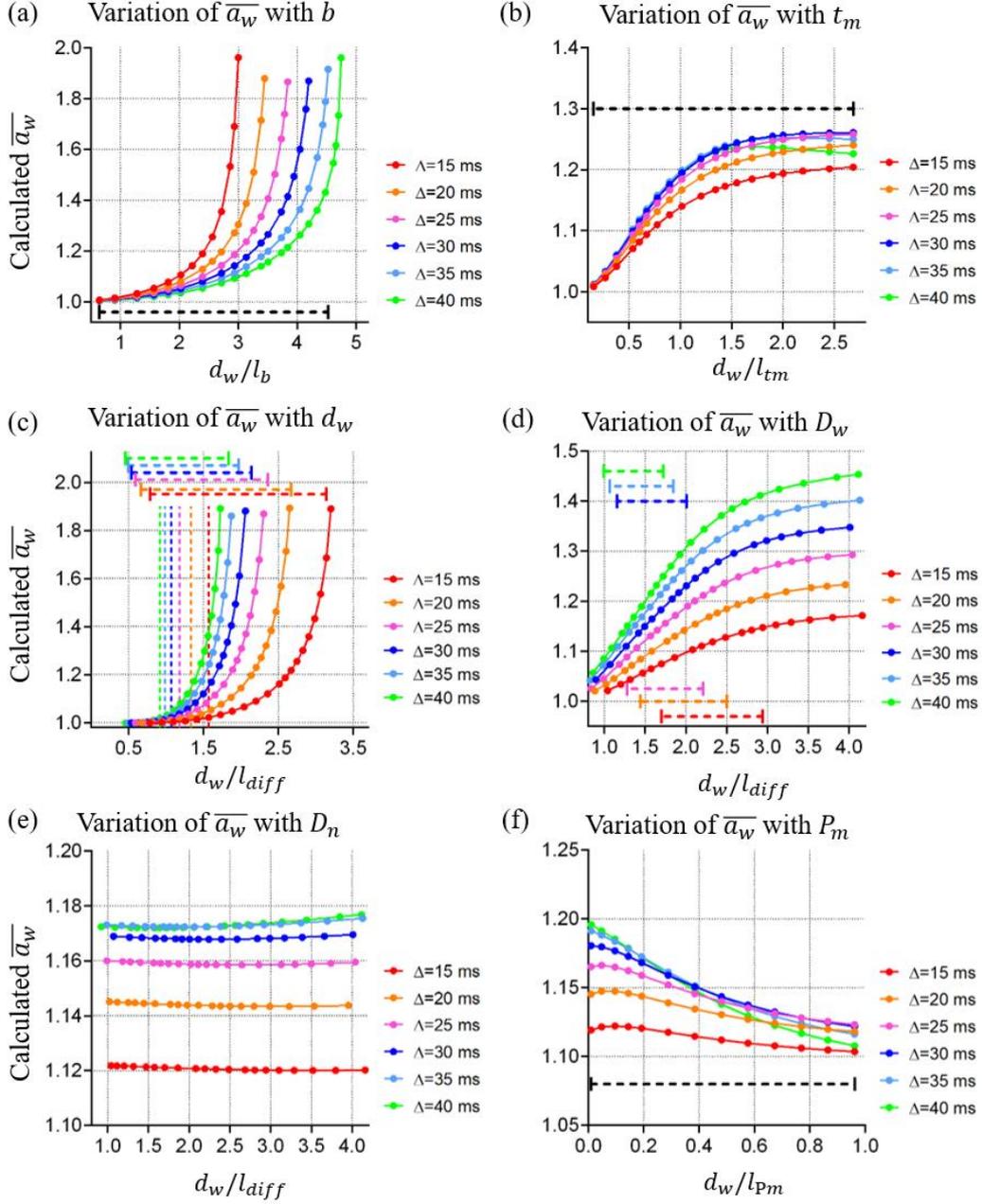

Figure 6. The calculated time-averaged ratio $\overline{a_w}$ as a function of the dimensionless ratio in the FEXI method. The basic settings in this group of simulations are: $\delta = 6$ ms, $t_m = 100$ ms, $g = 180$ mT/m, $d_w = 15$ μm, $d_n = 5$ μm, $D_w = D_n = 1.56$ μm²/ms, $P_m = 0.02$ μm/ms. (a). The variation of $\overline{a_w}$ with the $b$-value (adjust $b$ by changing $g$). The different colors of the dots and lines represent different $\Delta$ settings, and the other variables are the same as the basic settings (same in the subfigures (b)-(f)). The black horizontal dashed line indicates that $b$ varies from 0~5 ms/μm² (from left to right, same below). (b). The variation of $\overline{a_w}$ with $t_m$. The black horizontal dashed line indicates that $t_m$ varies from 3000~10 ms. (c). The variation of $\overline{a_w}$ with $d_w$. The colorful horizontal dashed line indicates that $d_w$ varies from 5~20 μm. (d). The variation of $\overline{a_w}$ with $D_w$. The colorful horizontal dashed line indicates that $D_w$ varies from 1~3μm²/ms. (e). The variation of $\overline{a_w}$ with $D_n$. (f). The variation of $\overline{a_w}$ with $P_m$. The black horizontal dashed line indicates that $P_m$ varies from 0.001~0.1μm/ms.

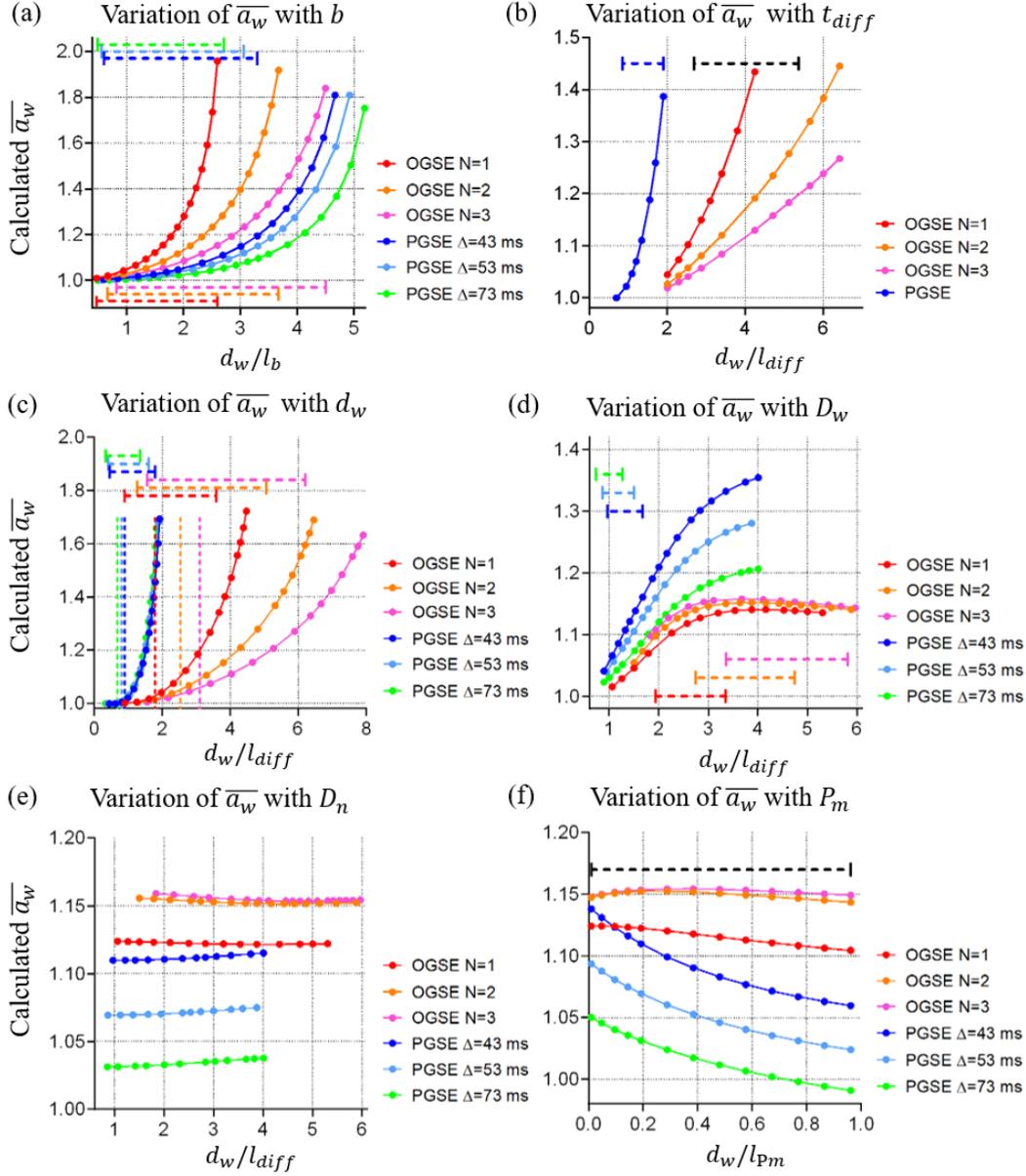

Figure 7. The calculated time-averaged ratio $\overline{a_w}$ as a function of the dimensionless ratio in the JOINT method. The basic settings in this group of simulations are: $\delta = 40\,\text{ms}$, $\Delta = 50\,\text{ms}$, $b = 1\,\text{ms}/\mu\text{m}^2$ for OGSE, and $\delta = 9\,\text{ms}$, $b = 2\,\text{ms}/\mu\text{m}^2$ for PGSE; in addition, $d_w = 15\,\mu\text{m}$, $d_n = 5\,\mu\text{m}$, $D_w = D_n = 1.56\,\mu\text{m}^2/\text{ms}$, $P_m = 0.02\,\mu\text{m}/\text{ms}$. (a). The variation of $\overline{a_w}$ with the $b$-value (adjust $b$ by changing $g$). The different colors of the dots and lines represent the OGSE sequences with different numbers of periods $N$ and the PGSE sequences with different $\Delta$ values, the other variables are the same as the basic settings (same in the subfigures (c)-(f)). The colorful horizontal dashed line indicates that $b$ varies from 0~3 ms/$\mu\text{m}^2$. (b). The variation of $\overline{a_w}$ with $t_{diff}$ (adjust $t_{diff}$ by changing $\delta$ in the OGSE or changing $\Delta$ in the PGSE). Here the different colors of the dots and lines represent the OGSE $N = 1,2,3$ and PGSE sequences, respectively, the other variables are the same as the basic settings.

The blue horizontal dashed line indicates that $t_{diff}$ varies from 20~100ms for PGSE, and the black horizontal dashed line indicates that $t_{diff}$ varies from 2.5~10ms for OGSE (c). The variation of $\overline{a_w}$ with $d_w$. The colorful horizontal dashed line indicates that $d_w$ varies from 5~20 μm. (d). The variation of $\overline{a_w}$ with $D_w$. The colorful horizontal dashed line indicates that $D_w$ varies from 1~3μm²/ms. (e). The variation of $\overline{a_w}$ with $D_n$. (f). The variation of $\overline{a_w}$ with $P_m$. The black horizontal dashed line indicates that $P_m$ varies from 0.001~0.1μm/ms.

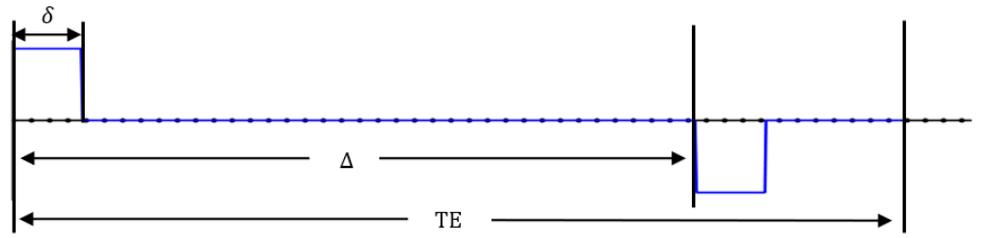
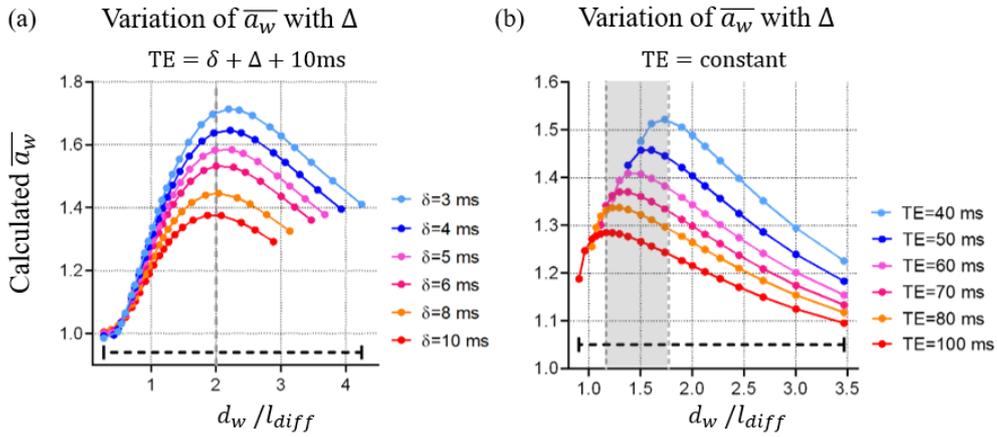

Figure 8. The calculated time-averaged ratio $\overline{a_w}$ as a function of the dimensionless ratio under a fixed $q = \gamma\delta g$. The basic settings in this group of simulations are: $\delta \cdot g = 1200\,\text{mT} \cdot \text{ms/m}$, $d_w = 15\,\mu\text{m}$, $d_n = 5\,\mu\text{m}$, $D_w = D_n = 1.56\,\mu\text{m}^2/\text{ms}$, $P_m = 0.02\,\mu\text{m/ms}$. (a). The overall duration TE was set to: $\text{TE} = \Delta + \delta + 10\,\text{ms}$. The different colors of the dots and lines represent different $\delta$ values. The other variables are the same as the basic settings. The black horizontal dashed line indicates that $\Delta$ varies from 1000~5 ms (from left to right, same below). (b). Here TE was set to the constant. The different colors of the dots and lines represent different TE values. $\delta = 6\,\text{ms}$ and $g = 200\,\text{mT/m}$, and the other variables are the same as the basic settings. The black horizontal dashed line indicates that $\Delta$ varies from 90~8 ms.

# Supplement Martials

Figures S1~S4 show the variations of $\overline{\alpha_{ex}}$ in the Kärger-model-based methods (CG, NEXI/SMEX, FEXI, and JOINT). Since almost all $\overline{\alpha_{ex}}$ values are close to 1, the variation curves are usually overlapped in the same figure.

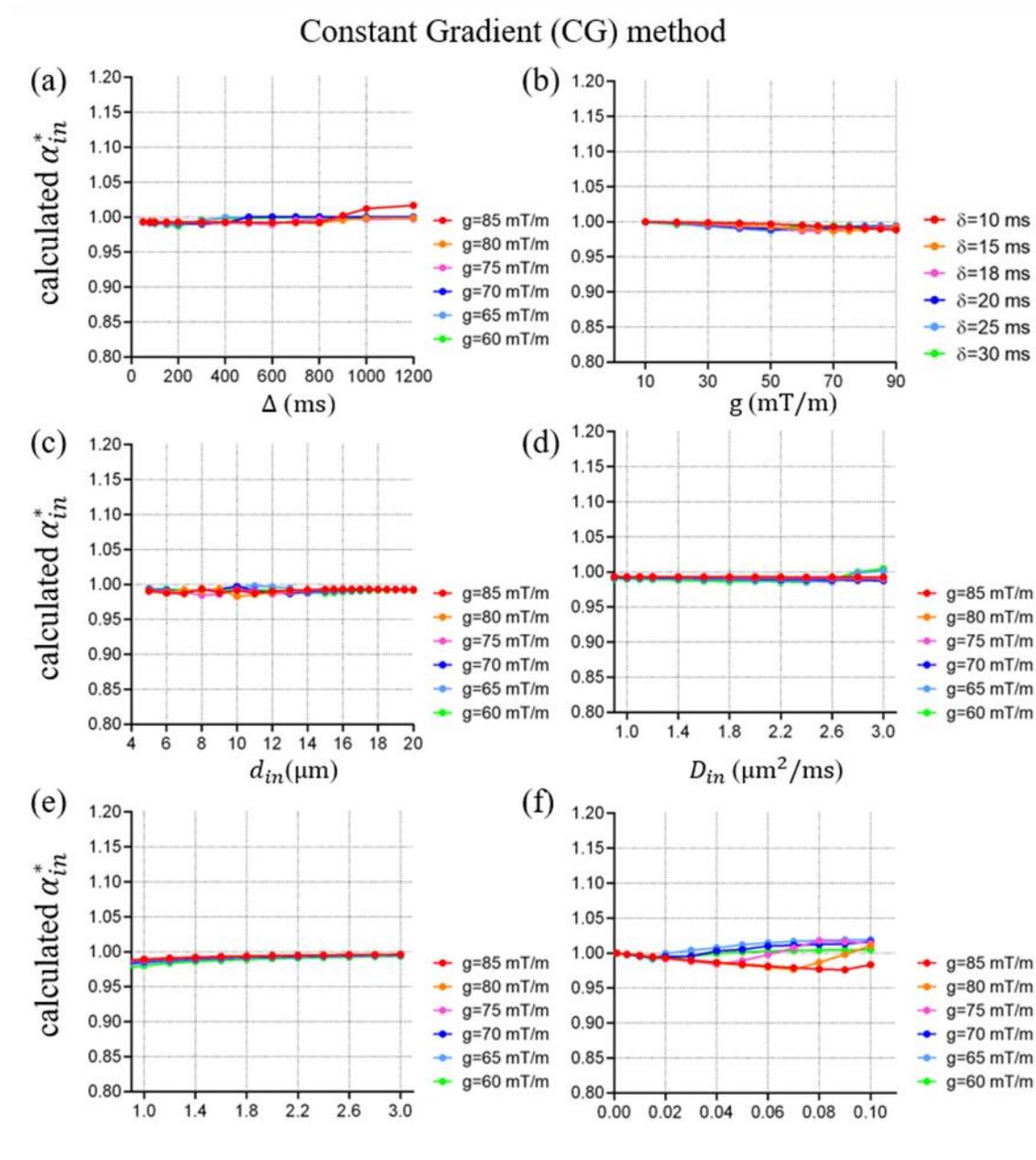

Figure S1. The calculated time-averaged ratio $\overline{\alpha_{ex}}$ as a function of each variable in the CG method.

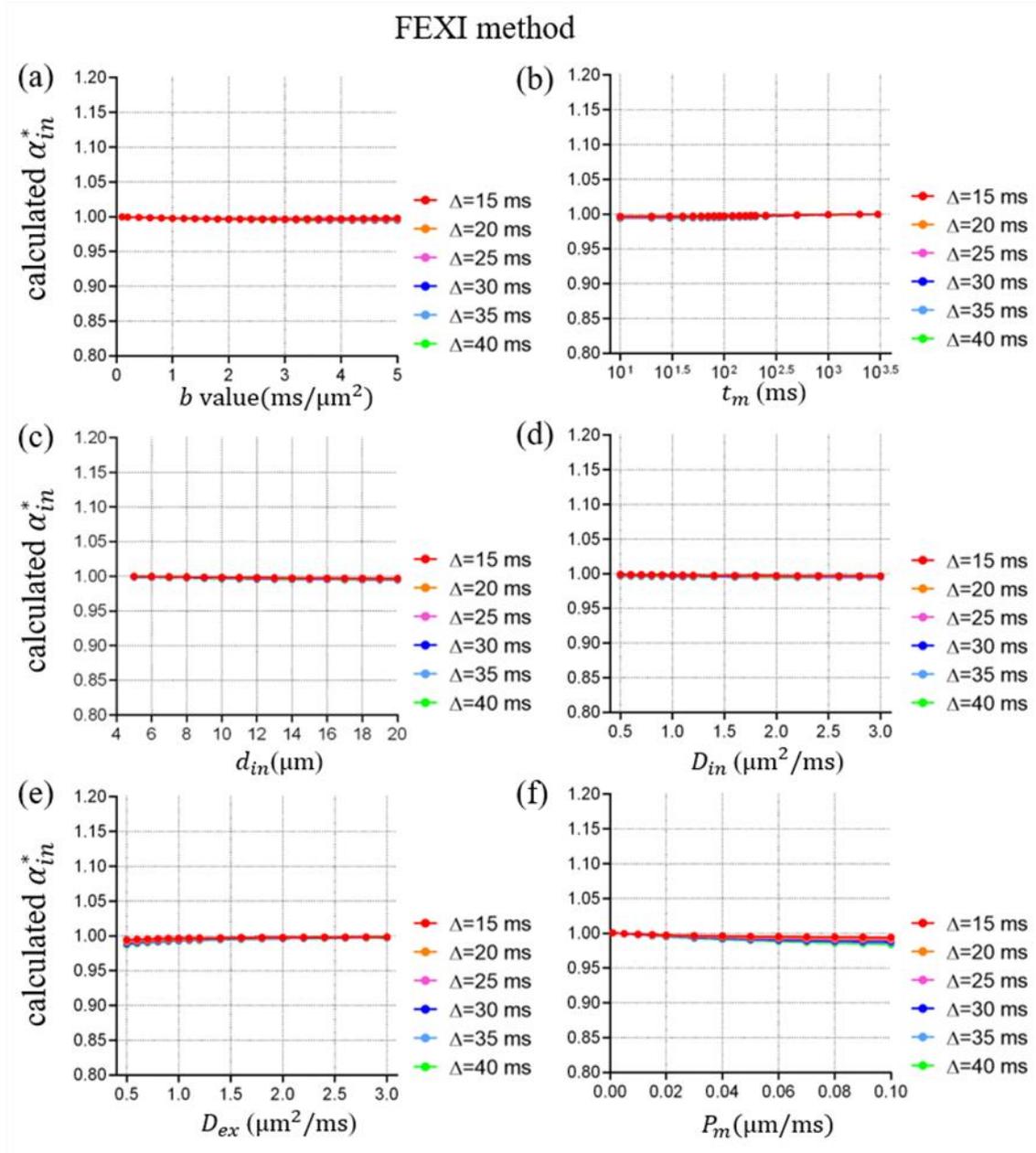

Figure S2. The calculated time-averaged ratio $\overline{\alpha_{ex}}$ as a function of each variable in the NEXI/SMEX method.

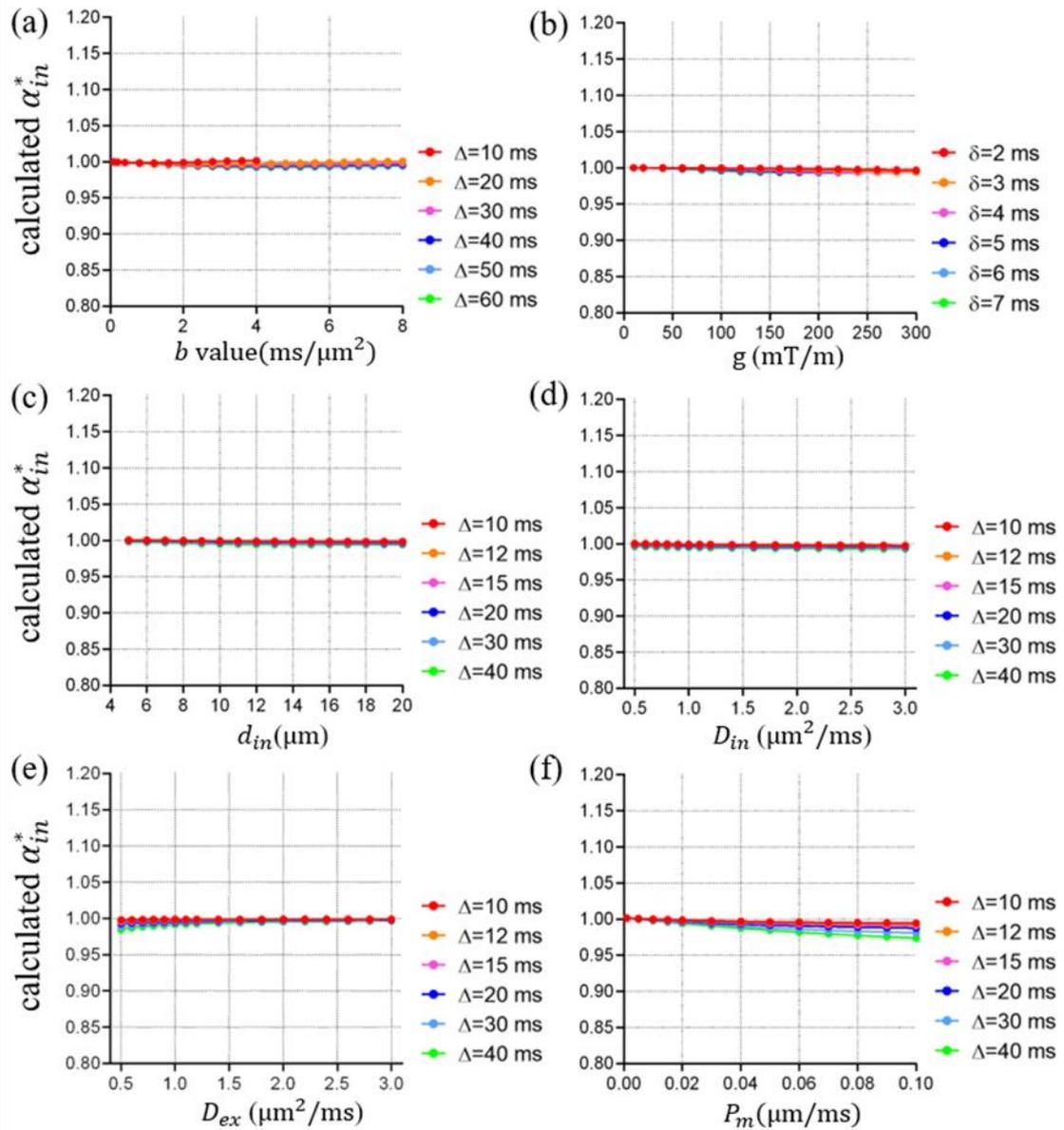

Figure S3. The calculated time-averaged ratio $\overline{\alpha_{ex}}$ as a function of each variable in the FEXI method.

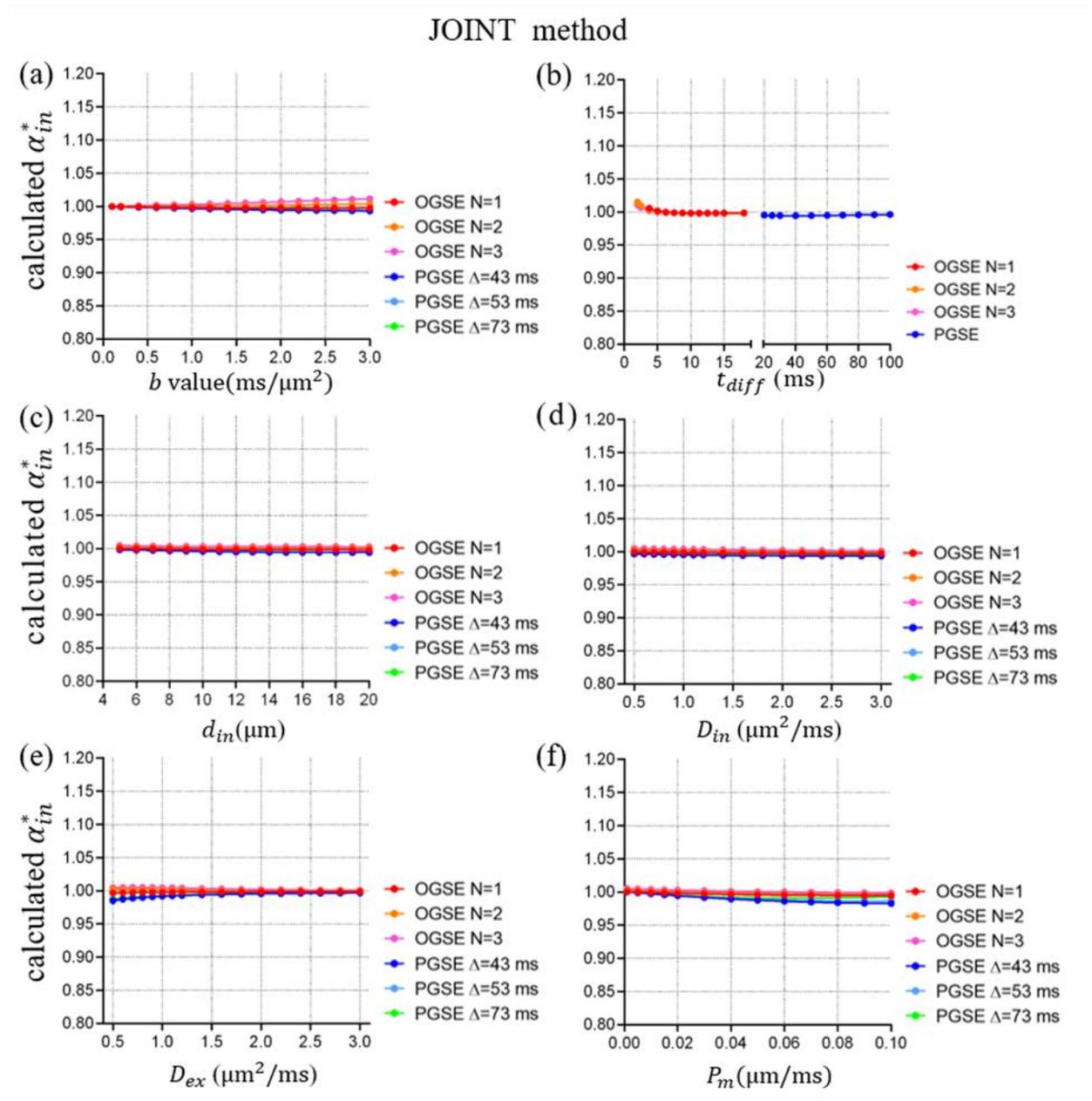

Figure S4. The calculated time-averaged ratio $\overline{\alpha_{ex}}$ as a function of each variable in the JOINT method.